\documentclass{article}
\usepackage{spconf,amsmath,graphicx,amsthm}

\usepackage{verbatim}
\usepackage{color}
\title{A  State Estimation and Malicious Attack Game \\in Multi-Sensor Dynamic Systems}
\name{Jingyang Lu and Ruixin Niu}
\address{Department of Electrical and Computer Engineering\\
Virginia Commonwealth University\\
Richmond, VA 23284, U.S.A.\\
\{luj2, rniu\}@vcu.edu}
\begin{document}
\newtheorem{theorem}{Theorem}
\newtheorem{proposition}{Proposition}
\theoremstyle{definition}
\newcommand{\beq}{\begin{equation}}
\newcommand{\eeq}{\end{equation}}
\newcommand{\beqa}{\begin{eqnarray}}
\newcommand{\eeqa}{\end{eqnarray}}
\newcommand{\nn}{\nonumber}
%\ninept
%
\maketitle
\begin{abstract}
In this paper, the problem of false information injection attack and defense on state estimation  in dynamic multi-sensor systems is investigated from a game theoretic perspective. The relationship between the Kalman filter  and the adversary can be regarded as a two-person zero-sum game. Under which condition both sides of the game will reach the Nash equilibrium is investigated in the  paper. The multi-sensor Kalman filter system and the adversary are supposed to be rational players. The Kalman filter and the adversary have to choose their respective subsets of sensors to perform system state estimation and false information injection. It is shown how both sides pick their strategies in order  to gain more and lose less. The optimal solutions are achieved by solving the minimax problem. Numerical results are also provided in order to illustrate the effectiveness of the derived optimal strategies.
\end{abstract}
\begin{keywords}
 Game theory, malicious attack, state estimation, Kalman filter, multi-sensor systems
\end{keywords}
\section{Introduction}
\label{sec:intro}
System state estimation in the presence of an adversary that injects false information into sensor readings is an important problem with wide application areas, such as target tracking with compromised sensors, secure monitoring of dynamic electric power systems and radar tracking and detection in the presence of jammers. This topic has attracted considerable attention and interest recently \cite{liu&etal:ccs09, jia&etal:icassp11,kosut&etal_icsgc10,jia&etal_pesgm12,Rahman&Mohsenian-Rad_globecomm12,kim&etal:jsac14,song&etal:sp12}. In \cite{liu&etal:ccs09}, the problem of how to take advantage of the power system configuration to introduce arbitrary bias to the system was investigated. In \cite{jia&etal:icassp11}, the authors showed the impact of malicious attacks on real-time electricity market and how the attackers can make profit by manipulating certain values of the measurements. The relationship between the attackers and the control center was discussed in \cite{kosut&etal_icsgc10}, where both the adversary's attack strategies and the control center's attack detection algorithms have been proposed. False data attacks on the electricity market have also been investigated in \cite{jia&etal_pesgm12} and  \cite{Rahman&Mohsenian-Rad_globecomm12}. In \cite{kim&etal:jsac14}, the data frame attack was formulated as a quadratically constrained quadratic program (QCQP). In \cite{song&etal:sp12}, the relationship between a target and a MIMO radar was characterized as a two-person zero-sum game. However, in the aforementioned publications, only the problem of {\it static} system state estimation has been considered.

 We are interested in {\it dynamic} system state estimation and in \cite{niu&huie_ssp12}, we have studied the impact of the injected biases on a Kalman filter (KF)'s estimation performance, showing that if the false information is injected at a single time, its impact  converges to zero as time goes on; if the false information is  injected into the system continuously, the estimation error tends to reach a steady state. In \cite{lu&niu:fusion14}, we have found the best strategy for the adversary to attack the Kalman filter system from the perspective of the trace of the mean squared error (MSE) matrix, and obtained  some close-form results. We have also studied how the attacker can maximize the determinant of the Kalman filter's estimation MSE matrix in  \cite{lu&niu:globalsip14}. Based on our previous work, in this paper our goal is to use game theory to investigate the relationship between the Kalman filter and the attacker.  The Kalman filter (the defender) and the attacker are supposed to be rational players. The trace of the state estimation MSE is used to construct the payoff matrix, and the problem can be characterized and solved as a minimax problem. Numerical results show the effectiveness of the optimal mixed defense strategy for the KF against the adversary's attacks.
\section{System Model}
\label{sec:sysmodel}
The discrete-time linear dynamic system can be described as 
\begin{equation}
	\label{eq:plant}
	\mathbf{x}_{k+1} = \mathbf{F}_{k} \mathbf{x}_{k} + \mathbf{G}_{k} \mathbf{u}_{k} + \mathbf{v}_{k}
\end{equation}
where ${\bf F}_{k}$ is the system state transition matrix, $\mathbf{x}_{k}$ is the system state vector at time $k$, $\mathbf{u}_{k}$ is a known input vector, $\mathbf{G}_{k}$ is the input gain matrix, and $\mathbf{v}_{k}$ is a zero-mean white Gaussian process noise with covariance matrix $E[\mathbf{v}_{k}\mathbf{v}_{k}^T] = \mathbf{Q}_{k}$. Let us assume that $M$ sensors are used by the linear system. The measurement at time $k$ collected by sensor $i$ is
\begin{equation}
	\label{eq:measure1}
	\mathbf{z}_{k,i} = \mathbf{H}_{k,i} \mathbf{x}_{k,i} + \mathbf{w}_{k,i}
\end{equation}
 with $\mathbf{H}_{k,i}$ being the measurement matrix, and $\mathbf{w}_{k,i}$ a zero-mean white Gaussian measurement noise with covariance matrix $ E[\mathbf{w}_{k,i} \mathbf{w}_{k,i}^T] = \mathbf{R}_{k,i}$, for $i=1,\cdots, M$. We further assume that the measurement noises are independent across sensors. The matrices $\mathbf{F}_{k}$, $\mathbf{G}_{k}$, $\mathbf{H}_{k,i}$, $\mathbf{Q}_{k}$, and $\mathbf{R}_{k,i}$ are assumed to be known with proper dimensions.
For such a linear and Gaussian dynamic system, the Kalman filter is the optimal state estimator. In this paper, we assume that a bias $\mathbf{b}_{k,i}$ is injected by the adversary into the measurement of the $i$th  sensor at time $k$ intentionally. Therefore, the measurement equation  becomes
\begin{equation}
	\label{eq:measure2}
	\mathbf{z}'_{k,i} = \mathbf{H}_{k,i} \mathbf{x}_{k} + \mathbf{w}_{k,i} +\mathbf{b}_{k,i}=\mathbf{z}_{k,i}+\mathbf{b}_{k,i}
\end{equation}  
where $\mathbf{z}'_{k,i}$ is the corrupted measurement, $\mathbf{b}_{k,i}$ is either an unknown constant or a random variable independent of $\{\mathbf{v}_{k,i}\}$ and $\{\mathbf{w}_{k,i}\}$. 

\section{Impact of False Information Injection}
\label{sec:impact}
Let us first assume that the adversary attacks the system by injecting false information into the sensors while the Kalman filter is unaware of such attacks. We start with the case where biases (${\bf b}_k$) are continuously injected into the system starting from a certain time $K$. Note that single injection is just a special case of continuous injection when ${\bf b}_k$ are set to be nonzero at time $K$ and zero otherwise.

In the continuous injection case, the Kalman filter' extra mean square error (EMSE), which is caused by the continuous bias injection alone, is derived in \cite{niu:summer_ext_faculty_12} and provided as follows.  
\begin{proposition}
\label{lem:mse_bias_continuous}
When the  bias sequence $\{{\bf b}_k\}$ is zero mean, random, and independent over time, the $EMSE$ at time $K+N$ due to the biases injected at and after time $K$, denoted as ${\bf A}_{K+N}$,  is $\mathbf{A}_{K+N}=\sum_{m=0}^{N}\mathbf{D}_{m}{\bf \Sigma}_{K+N-m} \mathbf{D}^T_{m}$, where $\mathbf{D}_{m}=\left(\prod_{i=0}^{m-1}\mathbf{B}_{K+N-i}\right)\mathbf{W}_{K+N-m}$, and $\mathbf{B}_{K}= \left(\mathbf{I}-\mathbf{W}_{K}\mathbf{H}_{K}\right) \mathbf{F}_{K-1}$. $\prod_{i=0}^{-1} \mathbf{B}_{K+N-i}={\bf I}$ is an identity matrix, $\mathbf{W}_{K}$ is the Kalman filter gain \cite{YBS:book}, and ${\bf \Sigma}_{K+N-m}$ is the covariance matrix of $\mathbf{b}_{K+N-m}$.
\end{proposition}

In \cite{lu&niu:fusion14}, we investigated the optimal attack strategy that an adversary can adopt to maximize the system estimator's estimation error. The problem can be formulated as a constrained optimization problem. Without loss of generality,  let us assume that the attacker is interested in maximizing the system state  estimation error at time $K$ right after a single false bias is injected at time $K$. In this case, we are interested in designing the injected random bias' covariance matrix such that 
\beqa 
	&&\max_{{\bf \Sigma}_K} \mathrm{Tr} \left[ {\bf P}_{K|K}+{\bf A}_K({\bf \Sigma}_K)\right]\nn\\ 
	&& s.t. \;\;\;\mathrm{Tr} ({\bf \Sigma}_K)=a^2
	\label{eq:max_trace}
\eeqa
where $a$ is a constraint on the power of the injected noise, $\mathrm{Tr}(\cdot)$ is the matrix trace operator, and ${\bf P}_{K|K}$ is the Kalman filter's state   covariance matrix at time $K$ in the absence of any false information. For both the cases where the attacker injects independent noises and dependent noises to position-only sensors in an object tracking system, we have derived the optimal strategies to  maximize the trace of the state estimation MSE matrix  as provided in the following two propositions \cite{lu&niu:fusion14}. 
\begin{proposition}
\label{pro:attack-ind}
For a system with $M$ sensors, if the adversary injects independent random noises, the best strategy  is to allocate all the power to the sensor with the smallest measurement noise variance.
\end{proposition}

\begin{proposition}
\label{pro:attack}
	For a system with $M$ sensors, the optimal strategy for the adversary is to inject dependent random noises with a pairwise correlation coefficient of $1$. The noise  power is allocated such that $\sigma_{b_i}=\frac{c_{i}a}{\sqrt{ \sum_{j=1}^M c^2_{j} }}, \;\;\;  i\in \{1,\cdots,M\}$, where $\sigma_{b_i}$ is the standard deviation (s.d.) of the noise injected to the $i$th sensor,  $
	c_{i}=\dfrac{1/{\sigma^2_{w_i}}}{\sum_{j=1}^M  \left(1/{\sigma^2_{w_j}}\right)}$, and $\sigma_{w_i}$ is the $i$th position-only sensor's measurement noise s.d. 
\end{proposition}
\section{A Target Tracking Example }

In this paper, we give a concrete target tracking example, and assume that the target moves in a one-dimensional space according to a discrete white noise acceleration model \cite{YBS:book}, which can still be described by the plant and measurement equations provided in (\ref{eq:plant}) and (\ref{eq:measure1}). 
In such a system, the state is defined as $\mathbf{x_k}=[\xi_k \;\; \dot{\xi}_k]^T$, where $\xi_k$ and $\dot{\xi}_k$ denote the target's position and velocity at time $k$ respectively.  The input $\mathbf{u}_k$ is a zero sequence. 
The state transition matrix is 
\begin{eqnarray}
\mathbf{F}_k=\left[\begin{array}{cc}
1 &\Delta\\
0 &1 	
\end{array}\right] \;\;\;\forall k
\label{eq:F}
\end{eqnarray}
where $\Delta$ is the sensor sampling  interval . The process noise is $\mathbf{v}_k=\mathbf{\Gamma} v_k$, where $v_k$ is a zero mean white acceleration noise, with variance $\sigma_{v}^{2}$, and the vector gain multiplying the scalar process noise is given by $ \mathbf{\Gamma} = \left[	\Delta^{2}/2\;\;\;  	\Delta \right]^T$.
The covariance matrix of the process noise is therefore $ \mathbf{Q}=\sigma^{2}_{v} \mathbf{\Gamma}  \mathbf{\Gamma}^T$. 
%The process noise has the autocorrelation function $E[v_kv_j]=\sigma_v^2 \delta_{kj}$, where 
%\beq 
%\delta_{kj}=\left \{ \begin{array} {ccc}
% 1& & \textrm{if} \;\;k=j\\
% 0& & \textrm{otherwise} 
% \end{array}
%\right.
%\eeq
%is the Kronecker delta function. 
The observation matrix in (\ref{eq:measure1}) is given as 
\begin{equation}
	{\bf H}_{k,i}=[1 \;\; 0]\;\;\forall k, \; i
\end{equation} 
Once the system model is known, it is straightforward for both the Kalman filter and the adversary to calculate the Kalman filter's state  covariance matrix ${\bf P}_{K|K}$ as in \cite{YBS:book}. Using Preposition 1, we can obtain the trace of the total state estimation MSE matrix:
\begin{equation}
	\textrm{Tr}(\textrm{MSE})=\textrm{Tr}({\bf P}_{K|K}+{\bf W}_K\boldsymbol{\Sigma}_K{\bf W}_K^{T})
\end{equation} 

\section{Noncooperative Two-Person Zero-Sum Game}

In a noncooperative two-person zero-sum game \cite{basar&olsder:book}, we assume that there are two players, referred to as Players 1 and  2, and an $m\times n $  payoff matrix ${\bf L}=\{l_{ij}\}$. Each entry of the matrix is an outcome of the game corresponding to a particular pair of decisions made by both players. Player 1 gets $m$ rows of the matrix as his/her strategy set, while for Player 2, the strategy set is the corresponding $n$ columns of the same matrix.
 
In our problem, suppose there are totally $M$ sensors, the Kalman filter and the adversary can choose any non-empty subsets of sensors to perform state estimation and attack respectively, which means $m=n=2^M-1$. ${\bf L}$ is a square matrix of the size $(2^M-1)\times (2^M-1)$. The payoff in the game between the Kalman filter system and the adversary will be the trace of the state estimation MSE matrix. For each set of sensors he/she chooses to attack,  the adversary is under a total injected noise power  constraint as  specified  in (\ref{eq:max_trace}). The Nash equilibrium between the Kalman filter and the adversary is achieved by solving the minimax optimization problem. %Our goal is to find the Nash equilibrium between the Kalman filter and the adversary.

Let \{row $i$, column $j$\} be a pair of strategies adopted by the players, and the corresponding outcome (payoff) be $l_{ij}$, which means that Player 1 should pay Player 2 the amount of $l_{ij}$. If $l_{i^*j}\le l_{i^*j^*}\le l_{ij^*}$, for all $i=1,\dots, m$ and all $j=1,\dots,n$, the pair \{$i^*,j^*$\} is  said to constitute a saddle-point equilibrium, and the game is said to have a saddle point in pure strategy. On the other hand,  if the pair of inequalities does not  exist, one can derive the mixed strategy to obtain the equilibrium. A mixed strategy is a probability distribution on the space of the player's pure strategies. A mixed strategy allows for a player to select a pure strategy randomly with a certain probability. In this case, the utility function $u$ is defined as
\begin{equation}
	u({\bf x},{\bf y})=\sum_{i=1}^{m}\sum_{j=1}^{n}x_il_{ij}y_j={\bf x}^{T} {\bf L} {\bf y}
\end{equation}
where ${\bf x}$ and ${\bf y}$ are the probability distribution vectors for the mixed strategies. Also, ${\bf x}\in X$, ${\bf y} \in Y$, where the set ${X}=\{{\bf x } \in R^m: {\bf x}\ge {\bf 0},\;\; \sum_{i=1}^{m}x_i=1\}$, and  ${Y}$ is defined in the same way. The Kalman filter playing as defender is trying to minimize the utility function $u({\bf x ,y})$ by choosing the best defending strategy, while the attacker wants to maximize the utility function  by choosing the best attack strategy. 
For the payoff matrix ${\bf L}$ of size $m\times n$, a vector of $\bf{x}^*$ is the best mixed strategy for the Kalman filter if 
\begin{equation}
	\overline{U}_{m}({\bf L})=\max_{{\bf y} \in Y}({\bf x}^*)^T{ {\bf L}}{\bf y} \le \max_{{\bf y} \in Y}{\bf x^T}{ {\bf L}}{\bf y}, {\bf x} \in X	
\end{equation}
The $\overline{U}_m({\bf L})$ is known as the average security level (loss ceiling) of the defender, the average security level (gain-floor) of the attacker $\underline{U}_m $ can also be defined as below,
\begin{eqnarray}
\underline{U}_m({\bf L})=\min_{{\bf x} \in X}{\bf x}^T{ {\bf L}}{\bf y}^* \ge \min_{{\bf x} \in X}{\bf x}^T{ {\bf L}}{\bf y}, {\bf y} \in Y
\end{eqnarray}

It always holds that $\overline{U}_m({ {\bf L}})= \underline{U}_m({ {\bf L}})$ for mixed strategies in noncooperative two-person zero-sum game. The saddle point in the mixed strategies is defined when the two bounds are equal to each other, which  can be found by solving the following  linear programming problem \cite{basar&olsder:book}:
\begin{eqnarray}
\label{eq:primal}
	\min_{{\bf x} \in X}&& b_u \\
	\text{s.t.}\;\;&&{\bf L}^T{\bf x} \le b_u{\bf 1}\nn\\
	&& {\bf x}^T{\bf 1}=1 \nn\\
	&& {\bf x} \ge 0\nn
\end{eqnarray}
where $b_u$ denotes a constant upper bound. For the attacker, the formula is the other way around,
\begin{eqnarray}
\label{eq:dual}
\max_{{\bf y} \in Y}&& b_l \\
\text{s.t.}\;\;&&{{\bf L}}{\bf y}\ge b_l{\bf 1}\nn\\
&& {\bf y}^T{\bf 1}=1  \nn\\
&& {\bf y} \ge 0\nn
\end{eqnarray}
where $b_l$ denotes a constant lower bound. From the formulation above, it is easy to see that (\ref{eq:dual}) is the dual form of the optimization problem (\ref{eq:primal}). The optimal function for the two problems are the same. Interested readers are referred to \cite{basar&olsder:book} for more details.

\section{Numerical Results}
\label{sec:numerical}
%Numerical results are presented to illustrate the relationship between the Kalman filter system and the attacker.
 In the example, for simplicity and ease of presentation, we assume that  there are three  sensors denoted as $\{z_1, z_2, z_3\}$ in the system having  independent measurement noises with   noise standard deviations    $\sigma_{w_1}=3$, $\sigma_{w_2}=4$, $\sigma_{w_3}=5$. The system process noise s.d. is  $\sigma_v=0.5$, sensors's sampling interval is $\Delta=1s$, and the system initial state ${\bf x}_0$ is assumed to follow a ${\cal N}(\hat{x}_{0|0},\; {\bf P}_{0|0})$ distribution, where $\hat{x}_{0|0}= [1\;\;1]^T$ and  
 \[ {\bf P}_{0|0}=\left[\begin{array}{cc}
 0.25 &0.25\\0.25 &0.5 \end{array}\right].\]
 
  The adversary  can choose any combination of sensors from the set  $ P_1=\{z_1,z_2,z_3, z_1z_2, z_1z_3, z_2z_3,z_1z_2z_3\}$ to attack with the power constraint of $\sum_{1}^{3}{\sigma_{b_i}}^2=100$, where $\sigma_{b_i}$  is the s.d. of the random noise injected to Sensor $i$.  Likewise,   the defender can choose any combination of sensors to perform  state estimation, and its  strategy set is the same: $P_2=P_1$. The game is played as below: if the defender uses data from Sensors $i$ and $j$ for state estimation, while the adversary attacks Sensors $i$ and $k$, then  system state estimation is affected by the false information from the $i$th sensor only. 
 
 In this game, the trace of the  state estimation MSE matrix is regarded as the payoff of the game. In the games of the independent  and dependent attacks, the system is attacked according to the strategies provided  in Propositions \ref{pro:attack-ind} and \ref{pro:attack} respectively. Let us assume that the  adversary attacks the sensors at time $k=100$, and  the payoff matrix is given in Tables \ref{table:independent} and \ref{table:dependent}. 
\begin{table}
\caption{Payoff  Matrix (Independent Case)}
\vspace{0.1in}
%\resizebox{0.8\columnwidth}{!}
{\footnotesize
	\begin{tabular}{|c|c|c|c|c|c|c|c|}
	\hline
	   $KF/At$       & $z_1$   &$ z_2$   & $z_3$  & $z_1z_2$ & $z_1z_3$ & $z_2z_3$ & $z_1z_2z_3$ \\ \hline
	    $z_1$       & 25.4 & 4.7  & 4.7 & 25.4  & 25.4  & 4.7   & 25.4     \\ 
	    $z_2$       & 7.2  & 23.5 & 7.2 & 7.2   & 7.2   & 23.5  & 7.2      \\ 
	    $z_3  $     & 10    & 10    & 23.6  & 10   & 10     & 10   & 10        \\ 
	    $z_1z_2$    & 13.5 & 6.6  & 3.4 & 13.5  & 13.5  & 6.6   & 13.5     \\ 
	    $z_1z_3$    & 16.4 & 3.8  & 5.4 & 16.4  & 16.4  & 3.8   & 16.4     \\ 
	    $z_2z_3$    & 5.0  & 12.4 & 8.0 & 5.0   & 5.0   & 12.4  & 5.0      \\ 
	    $z_1z_2z_3$ & 10.2 & 5.2  & 3.9 & 10.2  & 10.2  & 5.2   & 10.2     \\ \hline
	\end{tabular}
	}

\label{table:independent}
\end{table}
 \begin{table}
\caption{Payoff Matrix (Dependent Case)}
\vspace{0.1in}
 %\resizebox{0.9\columnwidth}{!}
{\footnotesize
 	\begin{tabular}{|c|c|c|c|c|c|c|c|}
 	\hline
 	   $KF/At$       & $z_1$   &$ z_2$   & $z_3$  & $z_1z_2$ & $z_1z_3$ & $z_2z_3$ & $z_1z_2z_3$ \\ \hline
 	    $z_1$       & 25.4 & 4.7  & 4.7 & 13.2  & 15.9  & 4.7   & 10.3     \\ 
 	    $z_2$       & 7.2  & 23.5 & 7.2 & 9.3   & 7.2   & 13.3  & 8.6      \\ 
 	    $z_3  $     & 10   & 10   & 23.6   & 10   & 11.0     & 12.1  & 10.5        \\ 
 	    $z_1z_2$    & 13.5 & 6.6  & 3.4 & 16.7  & 12.4  & 5.6   & 15.6     \\ 
 	    $z_1z_3$    & 16.4 & 3.8  & 5.4 & 15.0  & 18.1  & 4.2   & 15.0     \\ 
 	    $z_2z_3$    & 5.0  & 12.4 & 8.0 & 6.8   & 5.3   & 15.5  & 8.2      \\ 
 	    $z_1z_2z_3$ & 10.2 & 5.2  & 3.9 & 12.5  & 11.1  & 6.2   & 13.4     \\ \hline
 	\end{tabular}
 	}
 
 \label{table:dependent}
 \end{table}
From  Tables \ref{table:independent} and   \ref{table:dependent}, we can see that there is no pure strategy Nash Equilibrium. Instead, we  use mixed strategies to find the Nash Equilibrium. In order to obtain  the optimal probability distribution vector,  we solve the optimization problem formulated in (\ref{eq:primal}).
The  solution to (\ref{eq:primal}) is the optimal probability vector for the defender, and the   dual solution is the optimal mixed strategy for the attacker. The optimal  solutions for  independent- and dependent-attack cases are shown in Tables \ref{table:independent_strategy} and \ref{table:dependent_strategy} respectively.  

%For the independent case, we can see that for the Kalman filter, the values of last rows are greater than those of the other rows so that the probabilities are much larger than those choosing other strategies. In Fig. \ref{fig:strategy}, the straight lines are used to show the lose ceiling for the Kalman filter, each point is average of MSE for certain strategy the Kalman filter chooses to do the state estimation. For each case, four options are shown: best mixed strategy, the same mixed strategy to the adversary, the $3_{rd}$ pure strategy, playing each strategy equally. It turns out that optimal mixed strategy is the best choice for the Kalman filter.
%The optimal solution to (\ref{eq:opt}) is the values of the probability distribution vector for defender. Based on the strong duality theory, we can get the optimal dual solution which is the mixed strategy for the attacker. The final solutions under independent and dependent cases are shown in Table \ref{table:independent_strategy} and \ref{table:dependent_strategy}. 
 For the independent case, we can see from Table \ref{table:independent} that  $(6,6)$ and $(7,7)$ elements of the payoff matrix (${\bf L}$) are the smallest among the seven diagonal elements. This means that in the worst cases for the KF when its chosen sensor combination happens to be  the same as that being attacked by the adversary,  the strategies  $z_2z_3$  and $z_1z_2z_3$  will lead to the smallest state estimation MSEs.  In addition,  for the KF, the values of last two rows are relatively small. As a result, for the KF, the probabilities of  the last two strategies ($z_2z_3$ and $z_1z_2z_3$) are much larger  than those of other strategies, which are shown in Table \ref{table:independent_strategy}. 

In the dependent case, for the KF, the probabilities for the last two pure strategies ($z_2z_3$ and $z_1z_2z_3$) are relatively large as  shown in Table \ref{table:dependent_strategy}. This can be explained similarly as in the independent case.   In ${\bf L}$, the entries of the  rows corresponding to $z_3$, $z_1z_2$, and $z_1z_3$  are relatively large, so the KF assigns nearly zero probabilities to these three strategies. In  the first two rows of ${\bf L}$, even though the diagonal elements are large, the rest of the elements are relatively small, so strategies $z_1$ and $z_2$ are assigned significant probabilities for the KF as shown in Table \ref{table:dependent_strategy}.

%er than the those of the other rows.. This means that if the KF picks either $z_2z_3$ strategy or $z_1z_2z_3$ strategy, and even the attacker is attacking the same sensor combination as adopted by the KF,  the KF will suffer the smallest MSEs.  In addition,  for the Kalman filter, the values of last two rows are relatively  smaller than the those of the other rows. As a result, for the KF, the probabilities for the last two pure strategies ($z_2z_3$ and $z_1z_2z_3$) are much larger  than those  choosing other strategies, which is shown in Table \ref{table:independent_strategy}. 

%Assume that the adversary uses the best mixed strategy (the optimal dual solution to (\ref{eq:opt})) to attack the system, the average of $P^2_{11}$ is used as the effect of the false information. In the Fig. \ref{fig:strategy}, it is shown that choosing the optimal mixed strategy will help the Kalman filter get a better system state estimation.

  \begin{table}
	
 \caption{Optimal Strategy Probabilities (Independent Case)}
\vspace{0.1in}
 %\resizebox{0.9\columnwidth}{!}
{\footnotesize
  	\begin{tabular}{|c|c|c|c|c|c|c|c|}
		  	\hline
  	   $Player$       & $z_1$   &$ z_2$   & $z_3$  & $z_1z_2$ & $z_1z_3$ & $z_2z_3$ & $z_1z_2z_3$ \\ \hline
  	    KF       & 0.00 & 0.00  & 0.00 & 0.00  & 0.00  & 0.40   & 0.60     \\ 
  	    Attacker       & 0.14 & 0.22 & 0.00 & 0.14   & 0.14   & 0.22  & 0.24      \\ 
   \hline
  	\end{tabular}
  	}
  \label{table:independent_strategy}
  \end{table}
 \begin{table}
\small
\caption{Optimal Strategy Probabilities  (Dependent Case)}
%\resizebox{0.9\columnwidth}{!}
\vspace{0.1in}
{\footnotesize
 	\begin{tabular}{|c|c|c|c|c|c|c|c|}
	%small
 	\hline
 	   $Player$       & $z_1$   &$ z_2$   & $z_3$  & $z_1z_2$ & $z_1z_3$ & $z_2z_3$ & $z_1z_2z_3$ \\ \hline
 	    KF       & 0.16 & 0.14  & 0.00 & 0.00  & 0.00  & 0.37   & 0.33     \\ 
 	    Attacker       & 0.14 & 0.02 & 0.00 & 0.00   & 0.00   & 0.34  & 0.50      \\ 
  \hline
 	\end{tabular}
 	}
 \label{table:dependent_strategy}
 \end{table}

\begin{figure}[htb]
 \centering
 {\includegraphics[width=3.2in]{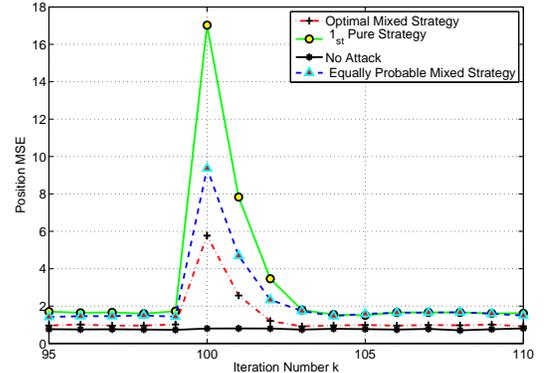}}
 \caption{Optimal Mixed Strategy vs. Other Options}
 \label{fig:strategy}
 \end{figure}

We also provide a simulation result to demonstrate the optimality of the derived strategy. In this example, four different scenarios are explored: 1) there is no attack; 2) the KF uses the optimal mixed strategy; 3) the KF uses a mixed strategy to pick each pure strategy with an equal probability $1/7$; 4) the KF always chooses the first pure strategy. In Scenarios 2)-4), the attacker injects false information according to his/her optimal mixed strategy to the sensors at time $k=100$.   The resulting position estimation MSEs are  plotted in Fig. \ref{fig:strategy}. It is clear that the optimal mixed strategy provides the best defense against the attacker, with the minimum increase in the MSE after the attack. 

\section{Conclusion}
In this paper, we investigated the relationship between the Kalman filter and the adversary in a two-person zero-sum game. The Kalman filter (defender) tries to achieve more accurate  system state estimation and avoid being attacked by the adversary. The adversary tries to mislead the Kalman filter as much as possible. Both sides of the game will reach a Nash Equilibrium through the mixed strategies. Using minimax techniques, we found  the mixed strategy saddle point in the game. In the future, we will put more practical constraints in our problem by letting both players in the game have limited information about the other player and introduce the detection mechanism to the Kalman filter system.

\bibliographystyle{IEEEbib}
\bibliography{Journal,Conf,Misc,Book}

\end{document}